\documentclass[prl,twocolumn,showpacs,preprintnumbers]{revtex4}
\usepackage{graphicx}
\usepackage{mathrsfs}
\usepackage{amsfonts}
\usepackage{amssymb}

\newcommand{\Tr}{\text{Tr}}
\newcommand{\ket}[1]{|#1\rangle}
\newcommand{\bra}[1]{\langle#1|}
\newcommand{\sz}{\sigma^z}
\newcommand{\sx}{\sigma^x}

\newcommand{\sigmam}{\sigma^-}
\newcommand{\sigmap}{\sigma^+}
\newcommand{\inner}[2]{\langle #1|#2\rangle} 
\newcommand{\ad}{a^{\dag}}
\newcommand{\p}[1]{\frac{\partial}{\partial t}#1} 

\begin{document}
\title{Entropy and specific heat for open systems in steady states}

\author{X. L. Huang, B. Cui, and  X. X. Yi}
\affiliation{School of Physics and Optoelectronic Technology\\
Dalian University of Technology, Dalian 116024 China}

\date{\today}

\begin{abstract}
The fundamental assumption of statistical mechanics is that  the
system is equally likely in any of the accessible microstates. Based
on this assumption,  the Boltzmann distribution is derived and the
full theory of statistical thermodynamics can be built. In this
paper, we show that the Boltzmann distribution in general can not
describe the steady state of open system. Based on the effective
Hamiltonian approach, we calculate the specific heat, the free
energy and the entropy for an open system in steady states. Examples
are illustrated and discussed.
\end{abstract}

\pacs{05.30.-d, 03.65.-w, 03.65.Yz} \maketitle

\section{introduction}
The fundamental postulate in statistical mechanics  is as follows:
For a conservative system the path of the system in phase space
passes through all points of the energy surface and in such a manner
that this surface is covered uniformly. This postulate (called
ergodic hypothesis)  indicates that a conservative system in
equilibrium does not have any preference for any of its available
microstates, i.e., given $\Omega$ microstates at a particular
energy, the probability of finding the system in a particular
microstate is $p = 1/\Omega$. By using the ergodic hypothesis, one
can conclude that for a system at equilibrium, the thermodynamic
state  which results from the largest number of microstates is the
most probable macrostate of the system. With these results, the
probability $p_i$ that a macroscopic system   in thermal equilibrium
with its environment in a given microstate with energy $E_i$ can be
derived, $p_i=exp(-\beta E_i)/\sum_j exp(-\beta E_j)$  and
$\beta=1/k_BT,$ which is  the so-called Boltzmann distribution.

It is believed that a generic system  that interacts with a generic
environment evolves into an equilibrium  described by the Boltzmann
distribution\cite{devi09}. Experience shows that this is true but a
detailed understanding of this process, which is crucial for a
rigorous justification of statistical physics and thermodynamics, is
lacking\cite{reimann08,goldstein06,popescu06,tasaki98,jensen85,bocchieri59}.
The key question is to what extent the evolution to equilibrium
depends on the details of the system-environment
coupling\cite{bertin09}. In fact, detailed analysis shows that it is
not always the case that the system evolves into the equilibrium
state described by Boltzmann distribution. Then two questions arise:
(1) Given a dynamical process, how far does the steady state differ
from the equilibrium state? (2)What is the free energy and specific
heat of such a system?

Here, we shall answer these questions by considering an open system
coupled to two independent environments at different temperatures.
The dynamics of the open system is assume to fulfill a master
equation in the Lindblad form, steady state is achieved by the
effective Hamiltonian approach. We find that the steady state is not
the thermal equilibrium state even if the environments have the same
temperature. Free energy, the specific heat and the entropy are
calculated and discussed.

\section{formalism}
\begin{figure}
\includegraphics*[width=0.6\columnwidth,
height=0.35\columnwidth]{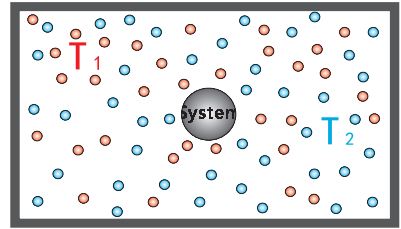} \caption{(Color online) A given
system immersed in two environments at temperature $T_1$ and $T_2$,
respectively. Eventually, the system arrives at a steady state
depending on the system-environment couplings and the temperature.}
\label{fig1}
\end{figure}

Imagine to prepare an open quantum system surrounded by two
independent environments at different temperatures $T_1$ and $T_2$,
respectively. See Fig.\ref{fig1}.
In the weak coupling limit and  under the Markov approximation, the
dynamics of the quantum system is govern by,
\begin{eqnarray}
i\p\rho&=&[H,~\rho]+\mathcal{L}_1\rho+\mathcal{L}_2\rho,
\label{mastereq}
\end{eqnarray}
where $\mathcal{L}_1\rho=\frac{i}2\sum_k\gamma_k(T_1)(2L_k\rho
L_k^{\dag}-L_k^{\dag}L_k\rho-\rho L_k^{\dag}L_k),
\mathcal{L}_2\rho=\frac{i}2\sum_k\Gamma_k(T_2)(2X_k\rho
X_k^{\dag}-X_k^{\dag}X_k\rho-\rho X_k^{\dag}X_k).$ $\rho(t)$
represents the reduced density matrix of the system, $H$ represents
the free Hamiltonian of the system and $L_k (L_k^{\dagger})$ and
$X_k (X_k^{\dagger})$ are operators of the system  through which the
system and its environments coupled. The master equation can be
solved by the effective Hamiltonian approach\cite{yi01}.

 The main idea of The effective Hamiltonian approach
can be outlined as follows. By introducing an ancilla, which has the
same dimension of Hilbert space as the system, we can map the system
density matrix $\rho(t)$ to a {\it wave function} of the composite
system (system + ancilla). A Schr\"odinger -like equation can be
derived from the master equation. The solution of the master
equation can be obtained by mapping the solution of the
Schr\"odinger -like equation back to the density matrix. Assume the
dimension of the Hilbert space for both the system and the ancilla
is $N$, and let $|E_n(0)\rangle$ and $|e_m(0)\rangle$ denote the
eigenstates for the system and the ancilla, respectively. The
mathematical representation of the above idea can be formulated  as
follows.  A wave function for the composite system in the
$N^2$-dimensional Hilbert space may be constructed as $
\rho(t)\rightarrow\ket{\Psi(t)}_{\rho}=
\sum_{m,n=1}^N\rho_{mn}(t)\ket{E_m(0)}\ket{e_n(0)},\label{map} $
where $\rho_{mn}(t)=\bra{E_m(0)}\rho(t)\ket{E_n(0)}$. Note that
$_{\rho}\inner{\Psi}{\Psi}_{\rho}=\Tr(\rho^2)\leq 1$, i.e. this pure
bipartite state is not normalized except when the state of the open
system is pure. With these definitions, the master equation
($\hbar=1$ hereafter) can be rewritten in a Schr\"odinger-like
equation\cite{yi01}
\begin{eqnarray}
i\frac{\partial}{\partial t} \ket{\Psi(t)}_{\rho}=\mathcal
{H}_{\text{eff}}\ket{\Psi(t)}_{\rho},\label{schrodinglike}
\end{eqnarray}
where $\mathcal {H}_{\text{eff}}$ is the so-called effective
Hamiltonian and is defined by
\begin{eqnarray}
&&\mathcal{H}_{\text{eff}}=\mathcal{H}-\mathcal{H}^A\nonumber\\
&&+i\sum_k \gamma_k(T_1)L_k^AL_k+i\sum_k\Gamma_k(T_2)X_k^AX_k,
\end{eqnarray}
where $ \mathcal{H}=H-\frac{i}2\sum_k\gamma_k(T_1)L_k^{\dag}L_k
-\frac{i}2\sum_k\Gamma_k(T_2)X_k^{\dag}X_k,$  $\mathcal {H}^A$,
$L^A_k$ and $X^A_k$ are  operators for the ancilla defined by ($O=L,
X, H$),
$\bra{e_m(0)}O^A\ket{e_n(0)}=\bra{E_n(0)}O^{\dag}\ket{E_m(0)}.$ By
this effective Hamiltonian approach, it is easy to prove that the
 steady state $\rho_S$ can be given by mapping the
eigenstate of $\mathcal{H}_{\text{eff}}$ with  zero eigenvalue.
Namely, calculating $\ket{R_0}$ by
\begin{eqnarray}
\mathcal{H}_{\text{eff}}\ket{R_0}=0,
\end{eqnarray}
we can obtain elements of the steady state density matrix, $
\rho_{S,mn}=\bra{E_m(0)}\rho_S\ket{E_n(0)}=\bra{E_m(0)}\inner{e_n(0)}{R_0}.
$ Given a steady state, the single-particle energy $U$ would be
equal to,
\begin{eqnarray}
U=\Tr(\rho_SH).
\end{eqnarray}
If a system consists of many non-interacting particles, the total
energy equals to the sum of the single-particle energy. The specific
heat for a single particle  now would be given by
\begin{eqnarray}
C_{T_i}=\frac{\partial U}{\partial T_i}.
\end{eqnarray}
Given the steady state density matrix $\rho_S$, von Neumann defined
the entropy as
\begin{eqnarray}
S=-\rho_S\ln\rho_S,
\end{eqnarray}
which is a proper extension of the Gibbs entropy  to the quantum
case.  We note that the entropy $S$ times the Boltzmann constant
$k_B$ ($k_B=1$ in this Letter) equals the thermodynamical entropy.
If the system is finite dimensional,  the entropy describes the
distance of the steady state from a pure state.

\section{examples}
\begin{figure}
\includegraphics*[width=0.8\columnwidth,height=0.6\columnwidth]{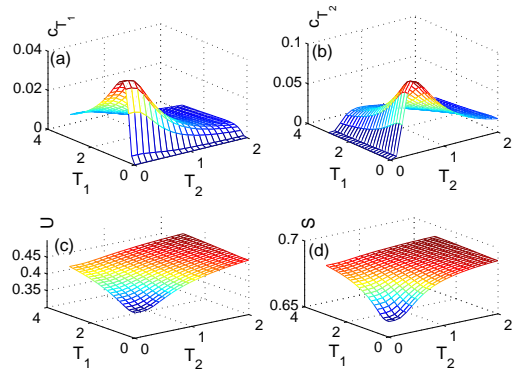}
\caption{(Color online)The specific heat, the free energy and the
entropy for the open two-level system as a function of temperature
(in units of $\frac{\hbar\Omega}{k_B}$). The parameters chosen are
$\gamma=0.2\Omega,$ $\Gamma=0.3\Omega.$ The energy was plotted in
units of $\hbar \Omega$. The units for the specific heat and entropy
were set accordingly.} \label{fig2}
\end{figure}

To illustrate the general formalism, we present here three examples.
In the first example, we consider a two-level system coupled to two
independent environments at different temperatures $T_1$ and $T_2$,
respectively. The dynamics is described by Eq.(\ref{mastereq}) with
$
\gamma_1(T_1)=\gamma\bar{n}_1,~~\gamma_2(T_1)=\gamma(\bar{n}_1+1),$
$ L_1=\sigmap,~~L_2=\sigmam, $
$\Gamma_1(T_2)=\Gamma(2\bar{n}_2+1),~~X_1=\sx $ and  $
\bar{n}_i=\frac{1}{e^{\beta_i\Omega}-1} (i=1,2).$ The system
Hamiltonian is specified to be $H=\frac{\Omega}2\sz.$ By the
effective Hamiltonian approach, we arrive at density matrix elements
of the steady state
$\frac{\rho_{11}}{\rho_{00}}=\frac{\gamma_1+\Gamma_1}{\gamma_2+\Gamma_1}$
with the trace preserving condition $\rho_{11}+\rho_{00}=1.$
Fig.\ref{fig2} depicts the specific heat $C_{T_1}, C_{T_2}$, the
free energy $U$ and the entropy $S$ as a function of temperature
$T_1$ and $T_2$. We note that for $T_1  \rightarrow 0$ and $T_1
\rightarrow \infty$ the specific heat $C_{T_1}$ tends to zero, the
population is then said to be frozen. For a fixed $T_1$ being of
order of $\Omega$, $C_{T_1}$ decreases as $T_2$ increases, and
$C_{T_1}$ tends to zero as $T_2 \rightarrow \infty$. $C_{T_2}$
behaves similarly.  The free energy $U$ and the entropy $S$ approach
constants with $T_1$ and $T_2$ tend to  infinity, confirming that
the population is frozen at sufficiently high temperatures. At low
temperature, $U$ and $S$ increase as the temperature increases,
indicating that the degree of mixture of the steady state grows with
the increasing of temperature.

\begin{figure}
\includegraphics*[width=0.8\columnwidth,height=0.6\columnwidth]{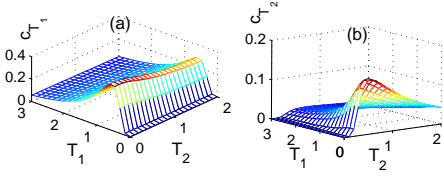}
\vspace{-2cm} \caption{(Color online) The specific heats $C_{T_1}$
and $C_{T_2}$ versus temperature $T_1$ and $T_2$ for two interacting
qubits dissipatively coupled to two independent environments.
$\Omega_1=\Omega_2, \gamma=0.2\Omega_1, J=0.2\Omega_1,
\Gamma=0.3\Omega_1$ are chosen for this plot.} \label{fig3}
\end{figure}

We take two coupled qubits subject to decoherence  as the second
example. The system Hamiltonian is,
$H=\Omega_1\ket{e}_1\bra{e}+\Omega_2\ket{e}_2\bra{e}+J\sx_1\sx_2. $
Suppose that the qubit 1 interacts with its environment at
temperature $T_1$ via $\sigma_1^{-}$ and $\sigma_1^{+}$, while the
qubit 2 through $\sigma_2^x$ couples to its environment at
temperature $T_2$. The Liouvillian superoperators are then,
\begin{eqnarray}
\mathcal{L}_1\rho=\frac{i}2\gamma(\bar{n}_1+1)(2\sigmam_1\rho\sigmap_1-
\rho\sigmap_1\sigmam_1-\sigmap_1\sigmam_1\rho)\nonumber\\
+\frac{i}2\gamma\bar{n}_1(2\sigmap_1\rho\sigmam_1-
\rho\sigmam_1\sigmap_1-\sigmam_1\sigmap_1\rho)\label{mstq1}
\end{eqnarray}
and
\begin{eqnarray}
\mathcal{L}_2\rho=\frac{i}2\Gamma(2\bar{n}_2+1)(2\sx_2\rho\sx_2-2\rho),\label{mstq2}
\end{eqnarray}
where
$\bar{n}_1=\frac{1}{e^{\beta_1\Omega_1}-1},~~\bar{n}_2=\frac1{e^{\beta_2\Omega_2}-1}.$
$C_{T_2} \rightarrow 0$ as $T_1$ and $T_2$ tend to $\infty$ as
Fig.\ref{fig3}(b) shows. For $C_{T_1}$, however, it approaches a
constant as $T_2\rightarrow \infty$ when $T_1$ takes a value of
order of $\Omega_i$  (Fig.\ref{fig3}(a)).
\begin{figure}
\includegraphics*[width=0.8\columnwidth,height=0.6\columnwidth]{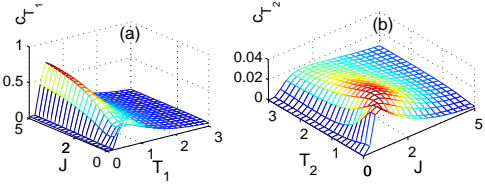}
\vspace{-2cm} \caption{(Color online) $C_{T_1}$ and $C_{T_2}$ as a
function of temperature and the coupling constant $J$. $T_2=1.5$ for
(a) and $T_1=1.5$ for (b). The other parameters are the same as in
Fig.\ref{fig3}.} \label{fig4}
\end{figure}
At $T_1=0,$ $C_{T_1}$ is always zero. For fixed $T_1$, $C_{T_1}$
tends to a constant with $J\rightarrow \infty,$ while $C_{T_2}$
always tends to zero as $J\rightarrow \infty,$  as Fig.\ref{fig4}
shows. Equilibrium statistical mechanics tells us that the
population of excited states (given by the Boltzmann distribution)
grows as the temperature increases. The populations obtained from
the steady state are different (see Fig.\ref{fig7}). E. g., the
population of the ground state increases as the temperature $T_2$
increases (Fig.\ref{fig7} (left)), whereas the population of the
second excited state (labeled by (3)) decreases as the temperature
increases. Similar observation can be found from
Fig.\ref{fig7}(right), the population of the first excited state
decreases as the temperature $T_1$ increases. We will quantify this
difference between the steady state and the equilibrium state by
fidelity later.
\begin{figure}
\includegraphics*[width=0.8\columnwidth,height=0.6\columnwidth]{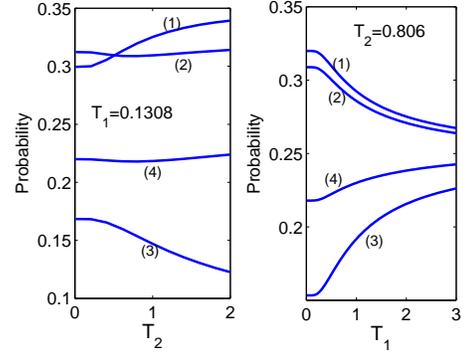}
\caption{Probability for finding the system in the eigenstates of
the Hamiltonian.  $\Omega_1=\Omega_2, \gamma=0.2\Omega_1,
J=0.2\Omega_1, \Gamma=0.3\Omega_1.$ (1)-(4) label the eigenstates by
the corresponding  eigenvalues  in increasing order, i.e., (1)
labels the lowest eigenstate, while (4) the highest eigenstate.}
\label{fig7}
\end{figure}

In the third example, we consider a damped harmonic oscillator. The
system Hamiltonian takes $ H=\omega \ad a. $ Consider a simple
system-bath (at temperature $T_1$) Hamiltonian of the form
$H_{int}=\sum_j g_j(b_ja^{\dagger}+b_j^{\dagger}a).$ The damping
rates follows by the standard procedure, $
\gamma_1(T_1)=\gamma\bar{n}_1, ~~L_1=\ad$, $
\gamma_2(T_1)=\gamma(\bar{n}_1+1),~~L_2=a. $ Suppose that the
harmonic oscillator interacts with the environment at temperature
$T_2$ through $(\ad+a)$, the corresponding damping rate is
$\Gamma_1(T_2)=\Gamma(2\bar{n}_2+1),$ and $ X_1=(\ad+a).$  All these
together give a master equation for the damped harmonic oscillator,
\begin{eqnarray}\label{msos}
i\p \rho&=&[H,~\rho]+\frac{i}2\gamma_1(2\ad\rho a-\rho
a\ad-a\ad\rho)\nonumber\\
&&+\frac{i}2\gamma_2(2a\rho\ad-\rho\ad a-\ad a\rho)\\
&&+\frac{i}2\Gamma_1[2(\ad{+}a)
\rho(\ad{+}a){-}(\ad{+}a)^2\rho{-}\rho(\ad{+}a)^2].\nonumber
\end{eqnarray}
\begin{figure}
\includegraphics*[width=0.8\columnwidth,height=0.6\columnwidth]{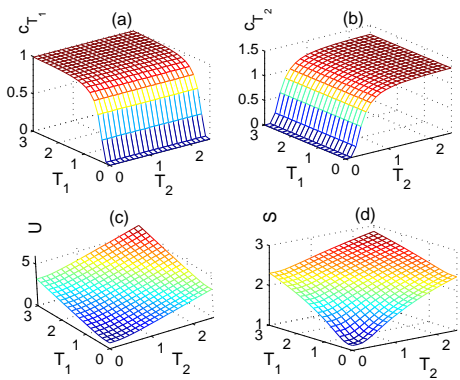}
\caption{(Color online) Illustration of $C_{T_1}, C_{T_2}, U$ and
$S$ as a function of temperature $T_1$ and $T_2$. In this plot, the
dimension cutoff is $N=100,$ and  $ \gamma=0.3\omega,
\Gamma=0.2\omega$. The energy is in units of $\hbar\omega$ and the
temperature is in units of $\frac{\hbar\omega}{k_B}.$} \label{fig5}
\end{figure}
Fig.\ref{fig5} shows the specific heat, the free energy, and the
entropy as a function of temperature. We note that the specific heat
$C_{T_1}$ ($C_{T_2}$) is vanishingly small for $T_1\rightarrow 0
(T_2\rightarrow 0),$ it rises rapidly when $T_1$ ($T_2$) is of order
of $\omega$ and approaches a limiting value, which depends on the
damping rates $\Gamma_1$ and $\gamma.$ Note that at sufficiently
high temperature, $C_{T_i}$ ($i=1,2$) are the same as that in
equilibrium statistical mechanics.

The results presented in the examples clearly show that $C_{T_i}
(i=1,2)$, $U$ and $S$ behaves different from those given by
equilibrium statistical mechanics. The differences result  from the
deviation of the steady state density matrix from the equilibrium
thermal state (Boltzmann distribution). We will use the fidelity to
quantify this deviation. Fidelity as a measure of distance  between
two states  is an important concept in quantum information
theory\cite{nielsen00}. The well-known quantum fidelity for two
general mixed states $\rho_1$ and $\rho_2$ is given by the Uhlmann's
fidelity\cite{bures69}
\begin{equation}
F(\rho_1,\rho_2)=\mbox{Tr}\sqrt{\sqrt{\rho_1}\rho_2\sqrt{\rho_1}},
\end{equation}
this fidelity possesses many advantages   such as concavity and
multiplicativity under tensor product and it satisfies all Josza's
four axioms\cite{jozsa94}.
\begin{figure}
\includegraphics*[width=0.8\columnwidth,height=0.6\columnwidth]{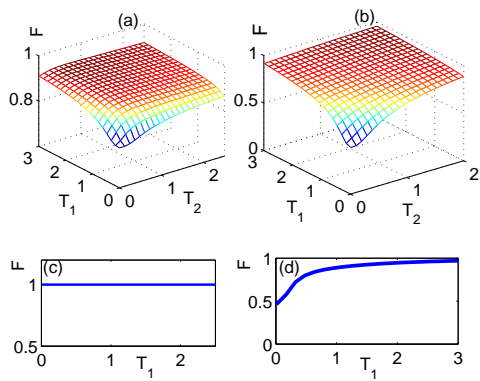}
\caption{(Color online) Distance between the steady states and
thermal states (Boltzmann distribution) measured by the fidelity as
a function of temperature. (a) and (c) are for the damped
oscillator, while (b) and (d) for the two coupled  qubits. The
steady state in (a) and (b) are obtained by solving the master
equations Eqs. (\ref{mstq1}), (\ref{mstq2}) and (\ref{msos}). The
steady states in (c) and (d) are for $\Gamma_1=0$ and $\Gamma=0,$
respectively. $\Omega_1=\Omega_2, J=0.2\Omega_1.$} \label{fig6}
\end{figure}
Fig.\ref{fig6} shows the fidelity between thermal states and the
steady states. For both the coupled qubits and the damped harmonic
oscillator, the fidelity arrives at its maximum when the
temperatures tend to infinity (see Fig.\ref{fig6}(a) and (b)). Note
that the steady states depends on how the system couples to its
environments. For example,  the master equation Eq.(\ref{msos}) with
$\Gamma_1=0$ can describe the thermalization of the oscillator, the
simulation shows that this is exactly  the case (Fig.\ref{fig6}(c)).
Differently,  for the coupled two qubits system, the steady state
given by the master equation (Eqs.(\ref{mstq1}) and (\ref{mstq2}))
with $\Gamma=0$ is not the equilibrium thermal state at low
temperature (see Fig.\ref{fig6} (d)).

One may have doubts about the realizability of the master equations
Eqs. (\ref{mastereq}), (\ref{mstq1}), (\ref{mstq2}) and
(\ref{msos}). The technology in engineered
reservoirs\cite{myatt00,cirone09} shows that there is no problem
to simulate an reservoir in which the system-environment coupling
and state of the environment are controllable.

In summary, we study the deviation of steady state from equilibrium
thermal state for an open system. The specific heat, free energy,
and entropy are calculated and discussed. This study applies to
several occasions, where a great many of physical phenomena of
interests concern collective behavior of an open system in steady
state. This work provides the exact solution for the nonequilibrium
distribution and statistical quantities  for steady states, thus
giving insight on how to build a statistical mechanics for open
systems.

This work is supported by NSF of China under grant Nos 10775023 and
10935010.

\end{document}